%Paper: hep-ph/9403308
%From: kim@cskim.yonsei.ac.kr (Choong Sun Kim)
%Date: Fri, 18 Mar 1994 15:09:46 +0900 (KST)

\input phyzzx
%\nopagenumbers
\hsize=6.4 true in \hoffset=0.2 true in
\vsize=9.0 true in \voffset=0.2 true in

\baselineskip= 16pt
\rightline{MAD/PH/817}
\rightline{YUMS 94--03 }
\rightline{SNUTP 94--11}
\rightline{(January 1994)}
\bigskip\bigskip
\bigskip\bigskip
\centerline{\bf Measuring $|V_{ub}|$ at future $B$--Factories\foot{\rm Talk
presented by C.S. Kim at {\it the 3'rd KEK Topical Conference on
CP--violation}, KEK, Japan, on
November 16--18.To be published by Nuclear Physics {\bf B} (Proc. Suppl.)
(1994).}}
\bigskip\bigskip
\bigskip\bigskip

\centerline{\bf
C.S. Kim$^a$\foot{\rm kim@cskim.yonsei.ac.kr (kim@165.132.21.18)},
Daesung Hwang$^a$\foot{\rm dshwang@cskim.yonsei.ac.kr},
Pyungwon Ko$^b$ and Wuk Namgung$^c$}

\medskip
\centerline{$^a$ Department of Physics, Yonsei University, Seoul 120--749,
KOREA}
\centerline{$^b$ Dept. of Physics, Univ. of Minnesota, Mineapolis,
MN. 55455, U.S.A.}
\centerline{$^c$ Department of Physics, Dongkook University, Seoul, KOREA}

\bigskip\bigskip
\centerline{\bf ABSTRACT}
\medskip

%\hsize=5.0 true in \hoffset=0.0 true in
%\hoffset=2.0 truein

\noindent
%We discuss about experimental problems and theoretical limitations on mesuring
%$V_{ub}$ of $B$--meson decays.
We calculate the so--called Fermi motion parameter
$p_{_F}$ of ACCMM model using the variational method in a potential model
approach. We also propose hadronic
invariant mass distribution as an alternative experimental observable to
measure $V_{ub}$  at future asymmetric $B$ factories.

\vfill\eject
%\end

%\hoffset=0.2 truein
\hsize=6.4 true in \hoffset=0. true in
\vsize=9.0 true in \voffset=0. true in
\bigskip\bigskip

\leftline{\bf 1. INTRODUCTION}
\medskip

In the standard $SU(2) \times U(1)$ gauge theory of Glashow, Salam and
Weinberg the fermion masses and hadronic flavor changing weak transitions
have a somewhat less secure role, since they require  a prior knowledge
of the mass generation mechanism. The simplest possibility to give mass
to the fermions in the theory makes use of Yukawa interactions involving
the doublet Higgs field. These interactions give rise to the
Cabibbo--Kobayashi--Maskawa (CKM) matrix: Quarks of different flavor are
mixed in the charged weak currents by means of an unitary matrix $V$.
However, both the electromagnetic current and the weak neutral current remain
flavor diagonal. Second order weak processes such as mixing and CP--violation
are even less secure theoretically, since they can be affected by both
beyond the Standard Model virtual contributions, as well as new physics
direct contributions. Our present understanding of CP--violation is based
on the three--family Kobayashi--Maskawa model [1] of quarks, some of whose
charged--current couplings have phases. Over the past decade, new data have
allowed one to refine our knowledge about parameters of this matrix $V$.

The CKM matrix element $V_{ub}$ (or ${|V_{ub}| / |V_{cb}|}$)
is measured through $B$-meson weak decay, but the determination of the value
is highly model dependent. One method is to investigate the end point region
of the lepton energy distribution in the inclusive semileptonic $B$-meson
decay: $B \rightarrow X_ul\nu$.
In Section 2, we assume the Gaussian ansatz of the so called ACCMM model [2]
and
determine the parameter $p_{_F}$ using the variational method in a potential
model approach. In Section 3,
we point out that directly measuring the invariant mass of final hadrons
in $B$--meson semileptonic decays offers an alternative way to select
$b \rightarrow u$ transitions [3] that is in principle more efficient than
selecting the upper end of the lepton energy spectrum.

%\vfill\eject
\medskip
\leftline{\bf 2. ACCMM MODEL AND THE PARAMETER $p_{_F}$}
\medskip

The simplest model for the semileptonic $B$-decay is the spectator model which
describes the decaying $b$-quark of the $B$-meson as a free particle.
The decay width with phase space and radiative corrections  can be written as
$$
\Gamma (b \rightarrow X_ql\nu) = | V_{bq}|^2 \Bigl({G_F{}^2 \, m_b^5 \over
192\pi^3}\Bigr) f\Bigl({m_q \over m_b}\Bigr) \Bigl[1 - {2\over 3}
{\alpha_s \over \pi} g\Bigl({m_q \over m_b}\Bigr)\Bigr]~. \eqno{(1)}
$$
QCD correction can be approximated [4] within
0.2\%,
$$g(\epsilon={m_q \over m_b}) \simeq (\pi^2 - 31/4) (1-\epsilon)^2 +3/2 \,.
\eqno(2)$$
Because of the $m_b^5$ factor in Eq.~(1), a small error on the $b$-quark mass
$m_b$ would be significantly amplified in the width. To take into account
the bound state effect and thus circumvent the $m_b^5$ factor problem,
a model (called the ACCMM model [2]) had been proposed.

The ACCMM model incorporates the bound state effect by treating the $b$-quark
as a virtual state particle, thus giving momentum dependence to the $b$-quark
mass. The momentum dependence of the virtual state $b$-quark mass is given by
$$
m_{b^*}^2 (\vec p\,) = m_B^2 + m_{sp}^2 - 2m_B \sqrt{\vec p\,^2 +
m_{sp}^2} \eqno{(3)}
$$
in the $B$-meson rest frame, where $m_{sp}$ is the spectator quark mass and
$m_B$ the $B$-meson mass. In this way the bound effect is introduced through
the spectator quark contribution to the decay width.

For the momentum distribution of the quark's Fermi motion inside the
$B$-meson, the ACCMM model simply assumes a Gaussian distribution,
$$
\phi (\vec p\,) = {4 \over \sqrt \pi p_{_F}^3} e^{-|\vec p |^2 / p_{_F}^2}
\eqno{(4)}
$$
with a free parameter $p_{_F}$. Thus the decay width is modified from the
spectator model to
$$
{d\Gamma_B \over dE_l} (p_{_F}, m_{sp}, m_q) = \int_0^{p_{max}} \!\!\! dp \;
p^2 \phi(\vec p\,) \, {d\Gamma_b \over dE_l} (m_q , m_b) \eqno{(5)}
$$
in the ACCMM model, where $\Gamma_b$ is given by eq.(1).

The model, therefore, introduces a new parameter, $p_{_F}$, for the momentum
measure of the Gaussian distribution, instead of the $b$-quark mass of the
spectator model. The new parameter $p_{_F}$ is regarded as free parameter,
and the value 0.3 is widely used in the literature without theoretical
justification. We assume  here the Gaussian $ansatz$ of the ACCMM model and
determine the parameter $p_{_F}$ using the variational method in a potential
model approach. By doing that we give some definite values to $p_{_F}$
in terms of the $b$-quark mass, $m_b$, and potential model parameters.
The Gaussian probability distribution of the momentum, Eq. (4), is
interpreted in our approach as the absolute square of the momentum
space wave function of the bound $B$-meson, {\it i.e.}
$$
\eqalign{\phi (\vec p\,) &= 4\pi | \chi (\vec p\,) |^2  \cr {\rm with}~~~~
\chi (\vec p\,) &= {1 \over (\sqrt\pi p_{_F})^{3/2}}
e^{-\vec p^2/ 2p_{_F}^2}~~. }
\eqno(6)
$$
The Fourier transform of the momentum
space wave function $\chi (\vec p\,)$ is regarded as the
position-space wave function $\psi (\vec r\,)$, which is itself Gaussian,
$$
\psi (\vec r\,) = \Bigl( {p_{_F} \over \sqrt\pi} \Bigr)^{3/2}
e^{-r^2 p_{_F}^2 /2}~. \eqno{(7)}$$

We will use the variational method with the Hamiltonian operator,
$$
H=\sqrt{\vec p\,^2 + m_{sp}^2} + \sqrt{\vec p\,^2+ m_b^2} + V(r)~~,
\eqno{(8)}
$$
and a trial function,
$$
\psi(\vec r\,) = { 1 \over (\sqrt\pi \mu )^{3/2}} e^{-\mu^2 r^2 /2}~~,
\eqno{(9)}
$$
where $\mu$ is a variational parameter.
The ground state is given by minimizing the expectation value of $H$,
$$
\eqalign{&\langle H \rangle = \langle \psi |H| \psi \rangle = E(\mu ) \cr
{\rm and}~~~~&{d \over d\mu} E(\mu ) = 0 \quad \rm at \quad \mu =
\bar \mu~~.\cr} \eqno{(10)}
$$
Then we approximate $m_B \sim E(\bar\mu)$ with $\bar\mu = p_{_F}$.
The $p_{_F}$ or $\mu$ of the Gaussian wave function corresponds to a measure
of the radius of the two-body bound state as can be seen from the
expectation values of $r$,
$$\langle r \rangle = {2 \over \sqrt\pi} {1 \over \mu}, \quad
\langle r^2 \rangle^{1 \over 2} = {3 \over 2} {1 \over \mu} \eqno{(11)}$$
or the most probable $r=1 / \bar\mu$

For the potential in the variational method, we use the linear plus coulomb
potential,
$$V(r) = -{\alpha_c \over r} + Kr~~. \eqno{(12)}$$
The best fit for the quark masses and the potential parameters,
$\alpha_c={4 \over 3}\alpha_s$ and $K$, had been determined by
Hagiwara {\it et. al.} [5],
$$
\eqalign{&\alpha_c = 0.47\; (\alpha_s = 0.35)~, \quad
K=0.19\;{\rm GeV^2}~~,  \cr
&m_b = 4.75\; {\rm GeV} \cr}
\eqno(13)
$$
for ($c \bar c$) and ($b \bar b$) bound states.
We will use these values in our analysis and also the value $\alpha_c
=0.32\; (\alpha_s=0.24=\alpha_s\; (m_b^2)\,)$ for comparison.

%With the reasonable assumption for the non--relativistic description of
%$b$-quark,
We use the relativistic kinematics only for the light $u$ or $d$ quark,
thus the Hamiltonian is written as
$$H \simeq M + {\vec p\,^2\over 2M} + \sqrt{\vec p\,^2+ m^2} + V(r)~,
\eqno{(14)}$$
where $M=m_b$ and  $m=m_{sp}$.
Trying to solve the eigenvalue equation of the differential operator (14)
may be faced with difficulty because of the square-root operator in $H$.
However, in our variational method, the expectation value of the $H$
can be calculated with either positron-space wave functions or
momentum-space wave functions,
$$
\langle H \rangle = \langle \psi (\vec r\,) | H |\psi (\vec r\,) \rangle
= \langle \chi (\vec r\,) | H |\chi (\vec r\,) \rangle~~.
\eqno{(15)}$$
Fortunately, our trial wave function is Gaussian both in position space
and in momentum space. Also the Gaussian function is a smooth function,
and the derivatives of any order are square integrable, that is, defined
on Hilbert space. Thus any power of the Laplacian operator $\nabla^2$ is
a hermitian operator at least under Gaussian functions. For more details,
see Ref. [6].

With  the input value of $m = 0.15$ GeV (as in the literature),
the calculated values are
$$\eqalign{&\bar \mu = 0.54, \qquad\bar E = 5.54 \qquad\qquad {\rm for}\;
\alpha_s=0.35~~, \cr {\rm and}~~~~
&\bar \mu = 0.49, \qquad\bar E = 5.63 \qquad\qquad {\rm for}\;
\alpha_s=0.24~~.}\eqno{(16)}$$
For comparison, we calculated $\langle H \rangle$ for the case of $m=0$ in
which the integral of the square root operator is exact,
$$
\eqalign{&\bar \mu = 0.53, \qquad\bar E = 5.52 \qquad\qquad {\rm for}\;
\alpha_s=0.35~~, \cr {\rm and}~~~~
&\bar \mu = 0.48, \qquad\bar E = 5.60 \qquad\qquad {\rm for}\;
\alpha_s=0.24~~.}\eqno{(17)}$$
The calculated values of the $B$-meson mass,  $\bar E$, are much larger than
the measured  value of 5.28.
The large values for the mass is originated mainly from the Hamiltonian
(14), which is flavor-degenerate for $B$ and $B^*$ (vector meson).
The difference between the pseudoscalar meson and the vector meson is the
chromomagnetic hyperfine splitting, which is given by Fermi-Breit as
$$V_s = {2 \over 3Mm} \; \vec s_1 \cdot \vec s_2 \,\nabla^2
(- {\alpha_c \over r})~~. \eqno{(18)}$$
And the expectation value of $V_s$ is given by
$$\eqalign{\langle V_s \rangle &= - {2 \over \sqrt\pi} \,
{\alpha_c \mu^3 \over Mm} \qquad{\rm for} \quad B~~,  \cr
&= {2 \over 3\sqrt\pi} \, {\alpha_c \mu^3 \over Mm} \qquad\; {\rm for}
\quad B^*~~.  }
\eqno(19)$$
Since $\langle V_s \rangle$ is proportional to $-\mu^3$ for $B$, $E(\mu)$ has
no minimum in $\mu$. Hence we can treat $\langle V_s \rangle$ only as a
perturbation for $B$-meson, resulting in
$$\eqalign{p_{_F} = 0.54, \qquad E_B = 5.52 \qquad\qquad
&{\rm for}\; \alpha_s=0.35~~, \cr {\rm and}~~~~
p_{_F} = 0.49, \qquad E_B = 5.56 \qquad\qquad
&{\rm for}\; \alpha_s=0.24~~.}
\eqno(20)$$
The perturbative result for $B^*$ is shown
$$\eqalign{p_{_F} = 0.54, \qquad E_{B^*} = 5.58 \qquad\qquad
&{\rm for}\; \alpha_s=0.35~~, \cr {\rm and}~~~~
p_{_F} = 0.49, \qquad E_{B^*} = 5.65 \qquad\qquad
&{\rm for}\; \alpha_s=0.24~~.}
\eqno(21)$$

The calculated values of the $B$-meson mass, 5.42 ($\alpha_s = 0.35$) and
5.56 ($\alpha_s = 0.24$) are in reasonable range compared to the
experimental value 5.28; the relative errors are 2.7\% and 5.3\%,
respectively.
But for $p_{_F}$, the calculated values, 0.54 ($\alpha_s = 0.35$) and  0.49
($\alpha_s = 0.24$), are much larger than the value 0.3 widely used in the
literature. For more details, see Ref. [6].

\medskip
\leftline{\bf 3. HADRONIC INVARIANT MASS DISTRIBUTION ON SEMILEPTONIC}
\leftline{\bf $B$--DECAY}
\medskip

The CKM matrix element $V_{ub}$ characterizing $b\to u$ quark transitions plays
an important role in the description of CP violation within the three-family
Standard Model, but is still not accurately known. The most direct way to
determine this parameter is through the study of $B$ meson semileptonic decays;
recent results from the CLEO [7]  and ARGUS [8] data on the
end-point region of the lepton spectrum have established that $V_{ub}$ is
indeed non-zero and have given an approximate value for its modulus.
The central problem in the extraction of $V_{ub}$ is
the separation of $b\to u$ events from the dominant $b\to c$ events. In
semi-leptonic $B$-meson decays, the usual approach  %\refmark{1,2}
 is to study the upper
end of the charged lepton spectrum, since the end-point region
$$
E_\ell > (m_B^2 - m_D^2 + m_\ell^2)/(2m_B)
\eqno(22)
$$
in the CM frame is inaccessible to $b\to c$ transitions and therefore selects
purely $b\to u$. However, only about 20\% of $b\to u$ transitions actually
lie in the region of Eq.~(22); it is therefore not a very efficient way to
select them. In situations of physical interest, the situation is even
somewhat worse.  For example, in $\Upsilon(4S) \to  B \bar B$ decay, each
$B$ meson has a small velocity in the $\Upsilon$ rest frame; the magnitude
$\beta$ of this velocity is known, but its direction is not. In this frame,
which is the laboratory frame when $\Upsilon$ is produced at a symmetric
$e^+e^-$ collider, the $b\to u$ selection region based on Eq.~(22) becomes
$$
E_\ell > \gamma(1+\beta)(m_B^2 - m_D^2)/(2m_B) ,
\eqno(23)
$$
%\vfill\eject
for the cases $\ell=e$ or $\mu$.     Here $\gamma = (1-\beta^2)^{-1/2} =
m_\Upsilon/(2m_B)$  and we neglect the lepton mass.  (At an asymmetric
collider, where $e^+$ and $e^-$ beams have different energies, it will be
necessary to boost lepton momenta from the laboratory frame to the
$\Upsilon$ rest frame before applying this cut.)  Equation~(23) accepts an even
smaller percentage of $b\to u$ decays than Eq.~(22), about 10\% in fact.

The essential physical idea behind Eqs.~(22)-(23) is that $b\to c$ transitions
leave at least one charm quark in the final state; hence for a general
semileptonic decay  $B \to  \ell + \nu + X $  the invariant mass $m_X$
of the final hadrons exceeds $m_D$ and this implies a kinematic bound on
$E_\ell$.
In Ref. 3, we gave this old idea a new twist. We first observe that
there is no unique connection
between $m_X$ and $E_\ell$, due to the presence of the neutrino, so the bound
on $E_\ell$ is not an efficient way of exploiting the bound on $m_X$.
We then observe a more efficient way to exploit the
latter bound is to measure $m_X$ itself and to select $b\to u$ transitions
by requiring
$$
m_X < m_D
\eqno(24)
$$
instead of Eqs.~(22)-(23). This condition is of course frame-independent.

The final hadronic invariant mass distribution depends both on the $c$-quark
energy distribution $d\Gamma(b \to c\ell\nu)/dE_c$ and on the Fermi momentum
distribution $\phi(p)$ which is normalized to $\int^\infty_0 dp \, \phi(p) =1$.
The lowest-order contribution to the $c$-quark energy
distribution is given by
$${d\Gamma^0(b \to c\ell\nu) \over dx_c} = {G^2_F m^5_b \over 96\, \pi^3}
|V_{cb}|^2 (x^2_c -4\epsilon^2)^{1/2} \left[ 3x_c (3-2x_c) + \epsilon^2 (3x_c-
4) \right] , \eqno(25)$$
where $x_c = 2(c\cdot b)/m^2_b=2E_c/m_b$ in the $b$ rest-frame, with
kinematical
range $2\epsilon \le x_c \le 1+\epsilon^2$.
When QCD radiative corrections are included, the real and virtual gluon
contributions must be subject to resolution smearing so that their singular
parts will cancel; this we approximate by absorbing real soft gluons into the
effective final $c$-quark and correcting $d\Gamma^0/dx_c$ by the factor
$g(\epsilon)$:
$${d\Gamma(b \to c\ell\nu) \over dx_c} \simeq {d\Gamma^0(b\to c\ell\nu) \over
dx_c } \left[ 1 -{2\alpha_s \over 3\pi}  g(\epsilon) \right] .  \eqno(26)$$
For each value of the Fermi momentum $p$ we calculate $d\Gamma/dE_c$ in the $b$
rest-frame (isotropic angular distribution here) and fold it with the spectator
energy-momentum vector to form the distribution $d\Gamma/dm_X$ with respect to
the invariant mass $m_X$ of the final charmed hadronic system,
$$m^2_X =(E_c + E_{\rm sp})^2 - ({\bf p}_c + {\bf p}_{\rm sp})^2. \eqno(27)$$
The spectator energy and momentum in the $b$ rest-frame are
$$\eqalign{E_{\rm sp} &= \left[ (p^2 +m^2_b)^{1/2} (p^2 +m_{\rm sp}^2)^{1/2}
+p^2 \right] /m_b \,, \cr
{\bf p}_{\rm sp} &= \left[ (p^2 +m^2_b)^{1/2} + (p^2 + m^2_{\rm sp})^{1/2}
\right] {\bf p} /m_b
\,, \cr} \eqno(28)$$
and $m_b$ is everywhere defined by Eq.~(3). The maximum and minimum
values of $m^2_X$ for given $p$ are
$$\eqalign{m^2_X(\max) &= m^2_c +m^2_{\rm sp} + m_b(E_{\rm sp} +p_{\rm sp})
+m^2_c(E_{\rm sp} -p_{\rm sp})/m_b \,, \crr
m^2_X(\min) &=
\cases{(m_c +m_{\rm sp})^2 \,, \qquad\qquad\qquad
 {\rm if} \ \  (m^2_b -m^2_c) E_{\rm sp}
\ge (m^2_b +m^2_c) p_{\rm sp}, &\cr \noalign{\vskip.1in}
m^2_c +m^2_{\rm sp} +m_b(E_{\rm sp} -p_{\rm sp}) + m^2_c(E_{\rm sp} +p_{\rm
sp})/m_b, \quad \ {\rm  otherwise}. \cr}
\cr}\eqno(29)$$
These relations show explicitly that small $p$ results in $m_X$ values close to
$(m_c +m_{\rm sp})$. The upper limit on $p$ for the decay to be possible (from
Eq.~(3) with $m_b >m_c$) is
$$p < \lambda^{1/2} (m^2_B, m^2_c, m^2_{\rm sp})/(2m_B) \,, \eqno(30)$$
where $\lambda(a,b,c) =a^2 +b^2 +c^2 -2ab -2bc -2ac$.

For $b \to u$ transitions the effect of individual resonances in $X$ quickly
disappears above the $\pi$ and $\rho$ region, and multiparticle jet-like
continuum final states should give the dominant contributions [9];
this makes
it reliable to calculate the $m_X$ distribution using the modified spectator
decay model.     Figure~1 gives the hadronic
invariant mass distribution from $B\to \ell \nu X$ semileptonic decays,
showing that more than
90\% of $b\to u$ decays lie within the region selected by Eq.~(24).  In this
illustration we use $m_B =5.273$~GeV, $E_B
=m_\Upsilon/2 =5.29$~GeV, $m_c=1.6$~GeV, $m_u = 0.1$~GeV,
$p_{_F}=0.3$~GeV,
including QCD corrections up to order $\alpha\alpha_s$ according to Ref. [10].
The figure is normalized for simplicity to the case $|V_{ub}/V_{cb}|=1$.
For more details, see Refs. [3,11].

In order to exploit Eq.~(24) instead, it is desirable to  isolate
uniquely the products of a single $B$ meson decay; to achieve this
it is generally necessary to reconstruct both $B$ decays in a given
event.  One of these decays can be semileptonic, since kinematic
constraints can often determine the missing neutrino four-momentum well
enough to reconstruct a peak at zero in the invariant square of this
four-momentum. Double-semileptonic decay events will not generally
reconstruct uniquely, however. To study semileptonic channels, we are
therefore concerned with those events (about 30\% of the total) where
one $B$ decays hadronically, one semileptonically with $\ell=e$ or $\mu$.
Of order 1\% of these have  $b\to u \ell\nu$  semileptonic transitions
(because of
the very small ratio $|V_{ub}/V_{cb}| \approx 0.1$).
About 10\% of the latter satisfy the criterion  $E_\ell > 2.5$~GeV  of
Eq.~(23).
With present data, it appears to be possible to reconstruct a few
percent of such events, but perhaps only about one percent without
ambiguity.

%\vfill\eject
%%Figure captions
\medskip
\noindent {\it Figure 1.}  Hadronic invariant mass distribution in $B \to \ell
\nu X$
semileptonic decays.
\vfill\eject

We note that there is a question of bias.  Some classes of final
states (e.g. those with low multiplicity, few neutrals) may be more
susceptible to a full and unambiguous reconstruction. Hence an
analysis that requires this reconstruction may be biassed. However
the use of topological information from microvertex detectors
should tend to reduce the bias, since vertex resolvability depends
largely on the proper time of the decay and its orientation relative
to the initial momentum (that are independent of the decay mode).
Also such a bias can be allowed for in the analysis, via suitable
modeling.
Finally there may be a background from
continuum events that accidentally fake the $\Upsilon$ events of interest.
This can be measured directly at energies close to the resonance.

\medskip
\leftline{\bf ACKNOWLEDGEMENTS}
\medskip

We would like to thank K. Abe, M. Tanaka and A.I. Vainshtein for usefull
discussions.
C.S.K. would like to thank V. Barger, A.D. Martin, R.J.N.
Phillips for the previous fruitful collaborations. This work was supported
in part by the Korean Science and Engineering
Foundation, in part by NON-DIRECT-RESEARCH-FUND, Korea Research Foundation
1993,
in part by the Center for Theoretical Physics, Seoul National University, and
in part by Yonsei University Faculty Research Grant.

\medskip
\leftline{\bf REFERENCES}
\medskip

\item{1.} M. Kobayashi and T. Maskawa,  Prog. Theo. Phys. 49 (1973) 652;
N. Cabibbo, Phy. Rev. Lett. 10 (1963) 531.

\item{2.} G.~Altarelli, N.~Cabbibo, G.~Corbo, L.~Maiani, and G.~Martinelli,
 Nucl. Phys. B208 (1982) 365.

\item{3.} V.~Barger, C.~S.~Kim, and R.~J.~N.~Phillips,  Phys. Lett. B235
(1990) 187; B251 (1990) 629.

\item{4.} C.S. Kim, and A.D. Martin,  Phys. Lett. B225  (1989) 186.

\item{5.} K. Hagiwara, A.D. Martin and A.W. Peacock, Z. Phy. C33 (1986) 135.

\item{6.} Daesung Hwang, C.S. Kim and Wuk Namgung, work in progress.

\item{7.} CLEO collaboration: R.~Fulton \etal,  Phys. Rev. Lett. 64
(1990) 16.

\item{8.} ARGUS collaboration: H.~Albrecht \etal,  Phys. Lett. B234
(1990) 409; B241  (1990) 278.

\item{9.} C.~Ramirez, J.~F.~Donoghue, and G.~Burdman, Phys. Rev. D41
(1990) 1496.

\item{10.} M.~Jezabek and J.~H.~K\" uhn,  Nucl. Phys. B320 (1989) 20.

\item{11.} C.S. Kim, Pyungwon Ko, Daesung Hwang and Wuk Namgung,
work in progress.

\end
\bye